\documentclass[conference]{IEEEtran}
\IEEEoverridecommandlockouts

\usepackage{cite}
\usepackage{amsmath,amssymb,amsfonts}
\usepackage{graphicx}
\usepackage{textcomp}
\usepackage{xcolor}
\usepackage{booktabs}
\usepackage{float}
\usepackage{placeins}
\usepackage{url}
\usepackage{hyperref}

\begin{document}

\title{CacheProbe: Auditing Prompt Cache Isolation in Gateway APIs}

\author{\IEEEauthorblockN{Ryan Fahey}
    \IEEEauthorblockA{\textit{Khoury College of Computer Sciences} \\
        \textit{Northeastern University}\\
        Boston, MA, USA \\
        fahey.rya@northeastern.edu}%
    \thanks{Accepted at SAGAI '26: Secure Agents for Generative Artificial Intelligence, co-located with the IEEE Symposium on Security and Privacy (S\&P) 2026, May 21, 2026, San Francisco, CA.}%
    \thanks{\copyright{} 2026 IEEE. Personal use of this material is permitted. Permission from IEEE must be obtained for all other uses, in any current or future media, including reprinting/republishing this material for advertising or promotional purposes, creating new collective works, for resale or redistribution to servers or lists, or reuse of any copyrighted component of this work in other works.}
}

\maketitle

\begin{abstract}
    Over the past year, prompt caching in Large Language Models (LLMs) has become increasingly more popular across inference APIs.
    Prompt caching helps save precious compute resources and speeds up response times by reusing parts of the KV cache of a specific prompt for another request.
    However, many implementations of prompt caching are not secure against timing attacks or even basic metadata disclosure.
    Gu et al.~\cite{DBLP:conf/icml/GuLKLH25} develop a method to audit prompt caching in LLMs.
    This paper investigates whether OpenRouter's API gateway architecture introduces prompt caching vulnerabilities that bypass provider-level prompt cache isolation guarantees.
    Most LLM inference providers implement per-account or per-organization prompt caching to prevent data leaks, but does routing through OpenRouter with shared organizational credentials inadvertently create global cache sharing across all OpenRouter users?
\end{abstract}


\section{Introduction}
\label{sec:introduction}

The rapid growth of LLMs, both in size and in usage, has caused the compute resources needed for inference to skyrocket.
As a result, prompt caching has become increasingly more popular across inference APIs.
Gim et al.~\cite{gim2024promptcache} describe prompt caching, a method in which key-value attention states can be cached and reused upon an initial request to an LLM.
Subsequent requests can use the same cached key-value attention states for portions of the prompt that are identical to the initial request.
This allows for only the attention computations of the unique tokens in the prompt to be performed, saving compute resources and speeding up response times.
Initial evaluations from the paper found that GPU inference was sped up by a factor of 1.5--10x while CPU inference was sped up by a factor of 20--70x.
Prompt caching allows providers to offer significant discounts on input tokens.
OpenAI, for example, offers an automatic 90\% discount on all cached input tokens in their most recent GPT-5 family of models on the standard tier.
Google's Gemini 2.5 and 3.0 family of models match this discount.

The authors of \cite{DBLP:conf/icml/GuLKLH25} discovered that via timing side-channels, it is possible to detect whether or not a provider implements prompt caching (which can leak model architecture details) as well as audit the contents of the cache.
If caches are shared across accounts, prompt caching timing signals can become a side-channel that leaks information about other users' prompts.
Since this work was published, most inference providers have implemented per-account or per-organization prompt caching to mitigate this risk.
However, API gateways like OpenRouter add an additional layer of abstraction to this issue.
If OpenRouter routes its users' requests to upstream providers using shared organizational credentials, then it is possible that OpenRouter is inadvertently creating global cache sharing across all OpenRouter users.

This paper investigates whether OpenRouter's API gateway architecture introduces prompt caching vulnerabilities that bypass provider-level prompt cache isolation guarantees, and it finds that it \textbf{does}.

\section{Background}
\label{sec:background}

\subsection{Prompt Caching in LLM APIs}

During inference, transformer-based LLMs compute key-value (KV) attention states for each token in the input.
For prompts of considerable length, this computation adds a significant amount of latency and incurs high computational costs.
Prompt caching largely fixes this issue by exploiting the fact that many requests (especially for multi-turn chat applications) share common prefixes and store and reuse those KV attention states.
Inference providers check if a prefix of a prompt matches a cached entry upon receiving a request.
If a match is found, the provider can reuse the cached KV attention states for the matching prefix and only compute the attention states for the unique tokens in the prompt.

\subsection{Timing Side-Channel Attacks}

The difference in latency between cache hits and misses creates a timing side-channel.
Attackers can measure the TTFT of a request to determine whether or not a given prefix exists in the cache.
By systematically probing with candidate prefixes, an attacker can reconstruct cached prompts token-by-token via binary search.
The authors of \cite{DBLP:conf/icml/GuLKLH25} formalize this attack and demonstrate its feasibility across major providers.

\subsection{OpenRouter Architecture}

OpenRouter is an API gateway that provides a unified interface for accessing inference APIs.
Rather than authenticating and sending requests to individual inference providers, users authenticate with their OpenRouter credentials and send their requests to OpenRouter.
OpenRouter routes the requests to the appropriate inference provider, authenticating with their own credentials.
From the provider's perspective, all the OpenRouter traffic is coming from a single organization, which raises questions on how the provider-level cache isolation is affected by this architecture.

\section{Threat Model}
\label{sec:threat-model}

\subsection{Adversary Goals and Capabilities}

We consider an adversary who is an authenticated user of OpenRouter, trying to detect or reconstruct prompts sent by other OpenRouter users.
The primary goal of the adversary is to determine whether or not a specific prompt prefix exists in the inference provider's cache, which indicates that another user recently submitted a prompt with that prefix.
If the adversary is able to detect prompt caching, it enables prompt extraction attacks as described in \cite{DBLP:conf/icml/GuLKLH25}, where an attacker can iteratively recover cached prompts token-by-token.

\subsection{OpenRouter Architecture Assumptions}

The adversary must possess a valid OpenRouter API key (with credits loaded) that will allow them to send arbitrary prompts to the target model while measuring the TTFT of each request.
Via OpenRouter's provider selection feature, the adversary can force their requests to be routed to a specific inference provider.
However, the adversary does \textbf{not} need any internal access to OpenRouter's infrastructure, the victim's credentials, or the victim's account.

\subsection{Detection Methods}

Our research exploits two signals to detect prompt caching.
The first is timing side-channel analysis: cache hits result in faster TTFT since the provider can skip KV attention computation for cached prefixes.
The second is metadata disclosure: some providers explicitly report how many tokens were served from cache via a \texttt{cached\_tokens} field in the API response.

\subsection{Attack Scenarios}

We evaluate six distinct scenarios to isolate the source of any cache sharing.

\begin{table}[h]
    \centering
    \footnotesize
    \setlength{\tabcolsep}{3pt}
    \caption{Attack Scenarios and Credential Configurations}
    \begin{tabular}{@{}llll@{}}
        \toprule
        \textbf{Scenario} & \textbf{Path} & \textbf{Accounts} & \textbf{Provider Creds} \\
        \midrule
        Direct Same       & Direct        & Same              & Same key                \\
        Direct Cross      & Direct        & Different         & Different keys          \\
        \midrule
        OR Same           & OpenRouter    & Same              & Shared (OR org)         \\
        OR Cross          & OpenRouter    & Different         & Shared (OR org)*        \\
        \midrule
        OR BYOK Same      & OpenRouter    & Same              & Same BYOK               \\
        OR BYOK Cross     & OpenRouter    & Different         & Different BYOKs         \\
        \bottomrule
    \end{tabular}
    \label{tab:scenarios}

    \vspace{0.3em}
    \scriptsize
    OR = OpenRouter. *Hypothesized shared provider credentials.
\end{table}

\section{Methodology}
\label{sec:methodology}

\subsection{Experimental Procedures}

During testing, we employed an interleaved sampling procedure that was designed to minimize any outside influence on the results, such as network latency or API load fluctuations.
For each sample pair in \texttt{num\_samples}, we first send a unique random prompt to establish a cache miss baseline, then have a simulated victim fill the cache with a different random prompt, and finally probe with a prefix-matching prompt to test for cache hits.
By interleaving miss and hit tests rather than batching them separately, it ensures that each pair is tested under similar conditions and any noise is duplicated within both distributions.

Prompts are generated as sequences of space-separated ASCII letters, since each letter preceded by a space constitutes a single token in most tokenizers.
We use a prompt length of 4,096 tokens with a 95\% prefix fraction; the attacker's probe shares the first 3,891 tokens with the victim's prompt but differs in the final 205 tokens.
This reflects real attack scenarios where an adversary attempts partial prefix reconstruction rather than exact prompt duplication.
A 500ms delay is inserted between requests to allow cache entries to persist while avoiding rate limiting.

\subsection{Test Scenarios}

Each of the six scenarios described in Section~\ref{sec:threat-model} is implemented by varying the API client configuration and credential assignment.
For direct scenarios, we instantiate OpenAI-compatible clients pointing to the provider's native endpoint with the appropriate API keys.
Same-account tests use identical credentials for both victim and attacker roles, while cross-account tests use distinct API keys obtained from separate account registrations.

For OpenRouter scenarios, requests are routed through OpenRouter's unified endpoint with provider selection forced via the \texttt{provider.order} parameter.
This ensures consistent routing to a single upstream provider rather than allowing OpenRouter's default load balancing across multiple providers.
In standard OpenRouter mode, both victim and attacker authenticate with different OpenRouter API keys, but OpenRouter authenticates to the upstream provider using its own shared organizational credentials.
In BYOK mode, users supply their own provider credentials via OpenRouter's credential management, meaning their requests authenticate to the provider under their personal accounts rather than OpenRouter's.

\subsection{Metrics and Statistical Tests}

We employ two complementary detection methods: timing-based statistical analysis and metadata disclosure analysis.
For timing analysis, we use the two-sample Kolmogorov-Smirnov test to compare the TTFT distributions of cache hit and miss samples.
Following \cite{DBLP:conf/icml/GuLKLH25}, we use a stringent significance threshold of $p < 10^{-8}$ to minimize false positives given the inherent noise in network timing measurements.

For metadata analysis, we examine the \texttt{cached\_tokens} field returned in API response usage metadata.
Some providers explicitly report how many input tokens were served from cache, providing ground-truth evidence of cache utilization independent of timing signals.
We consider metadata-based detection positive when more than 50\% of cache-primed requests report cached tokens exceeding 90\% of the expected prefix length.

Our combined detection criterion considers caching detected if \textit{either} method yields a positive result.
This conservative approach accounts for scenarios where timing signals may be obscured by network noise but metadata disclosure is present, or vice versa.

\section{Experimental Setup}
\label{sec:setup}

\subsection{Provider Selection}

We selected three inference providers for evaluation: OpenAI, Groq, and Fireworks.
These providers were chosen because they each implement organization-level prompt caching with documented isolation guarantees and are available through OpenRouter's unified API.

For each provider, we tested a single model: \texttt{gpt-4o-mini} for OpenAI, \texttt{openai/gpt-oss-20b} for Groq, and \texttt{qwen/qwen3-8b} for Fireworks.
Model selection was constrained by OpenRouter compatibility requirements; notably, we chose \texttt{gpt-4o-mini} over newer models like \texttt{gpt-4.1-nano} because OpenRouter's routing layer does not permit \texttt{max\_tokens=1} for certain models, which complicates precise Time to First Token (TTFT) measurements.

\subsection{Implementation}

Our testing harness is implemented in Python using the OpenAI SDK.
All experiments were conducted on an Ubuntu virtual machine hosted on Oracle Cloud in the US-East (Ashburn, Virginia) region to ensure consistent network conditions and eliminate variability from residential network connections.
Experiments were conducted between November 12 and December 8, 2025.

For each provider, we collected 250 sample pairs per scenario using the interleaved procedure described in Section~\ref{sec:methodology}.
Prompts consisted of 4,096 tokens with a 95\% prefix fraction, and we inserted a 500ms delay between requests to allow cache entries to persist while avoiding rate limiting.
The complete test matrix spans six scenarios per provider, resulting in 18 distinct experimental configurations and approximately 9,000 total API requests.

\section{Results}
\label{sec:results}

The following table summarizes our detection results across all provider-scenario combinations.
We report the Kolmogorov-Smirnov statistic and p-value for timing analysis, mean TTFT for cache miss and hit conditions, and whether caching was detected via statistical analysis (p $<$ $10^{-8}$), metadata disclosure, or either method.

\begin{table}[htbp]
    \centering
    \caption{Detection results across all provider-scenario combinations.}
    \label{tab:results}
    \footnotesize
    \setlength{\tabcolsep}{4pt}
    \begin{tabular}{@{}llccccc@{}}
        \toprule
        Provider  & Acct  & p-value & Miss & Hit  & Timing     & Meta                 \\
        \midrule
        \multicolumn{7}{l}{\textit{Direct API}}                                       \\
        Groq      & Same  & 3.8e-11 & .460 & .766 & \checkmark & \checkmark           \\
        Groq      & Cross & 8.3e-03 & .432 & .292 & --         & --                   \\
        Fireworks & Same  & 3.8e-11 & .445 & .403 & \checkmark & --                   \\
        Fireworks & Cross & 6.1e-01 & .374 & .379 & --         & --                   \\
        OpenAI    & Same  & 1.3e-01 & .825 & .784 & --         & \checkmark           \\
        OpenAI    & Cross & 6.9e-01 & .870 & .900 & --         & --                   \\
        \midrule
        \multicolumn{7}{l}{\textit{OpenRouter Default}}                               \\
        Groq      & Same  & 2.4e-03 & .413 & .351 & --         & \checkmark           \\
        Groq      & Cross & 1.7e-03 & .396 & .365 & --         & \checkmark           \\
        Fireworks & Same  & 3.3e-13 & .468 & .419 & \checkmark & --                   \\
        Fireworks & Cross & 4.1e-15 & .428 & .386 & \checkmark & --                   \\
        OpenAI    & Same  & 4.5e-03 & .940 & .572 & --         & \checkmark           \\
        OpenAI    & Cross & 6.1e-01 & .574 & .858 & --         & \checkmark$^\dagger$ \\
        \midrule
        \multicolumn{7}{l}{\textit{OpenRouter BYOK}}                                  \\
        Groq      & Same  & 4.0e-01 & .416 & .405 & --         & \checkmark           \\
        Groq      & Cross & 6.9e-01 & .429 & .428 & --         & --                   \\
        Fireworks & Same  & 3.7e-47 & .332 & .281 & \checkmark & \checkmark           \\
        Fireworks & Cross & 6.1e-01 & .475 & .513 & --         & --                   \\
        OpenAI    & Same  & 2.9e-01 & 1.77 & 1.36 & --         & \checkmark           \\
        OpenAI    & Cross & 1.6e-01 & 1.58 & 1.27 & --         & --                   \\
        \bottomrule
    \end{tabular}

    \vspace{0.3em}
    \footnotesize
    Timing: $p<10^{-8}$. $^\dagger$OpenAI cross-account: 4.8\% cache hit rate.
\end{table}

\FloatBarrier

\subsection{Direct API Baseline}

When accessing providers directly, all three providers demonstrate proper cross-account cache isolation.
Same-account tests confirm that prompt caching is functional: Groq and Fireworks both show statistically significant timing differences (p $<$ $10^{-8}$), while OpenAI reports cached tokens via metadata disclosure.
Notably, Groq's same-account cache hits exhibited higher TTFT than misses despite functional caching confirmed via metadata.
Cross-account tests show no evidence of cache sharing for any provider, with p-values ranging from 0.13 to 0.69 and no cross-account cached token reports.
This establishes that the providers' native isolation mechanisms function correctly.

\subsection{OpenRouter Default Mode}

When routing through OpenRouter with default credentials, we observe cache sharing across accounts for all three providers.

\textbf{Groq:} While timing analysis does not reach statistical significance (p = 0.002), metadata disclosure reveals 100\% cache hit rates in both same-account and cross-account OpenRouter scenarios.
Of the samples that reported cached token counts, all exceeded the 90\% prefix threshold in cross-account tests, indicating that prompts submitted by one OpenRouter user are being served from cache to other users.

\textbf{Fireworks:} Timing analysis provides strong evidence of cross-account cache sharing, with the OpenRouter cross-account scenario yielding p = $4.08 \times 10^{-15}$.
This represents the most statistically significant cache sharing detection in our study.
Fireworks does not report cached token metadata, so timing analysis is the only detection method available.

\textbf{OpenAI:} Metadata analysis reveals minor cache leakage: 4.8\% of cross-account requests report cached tokens exceeding the prefix threshold.
While substantially lower than Groq's 100\% rate, any cross-account cache hits with randomly-generated prompts indicate that caches are not fully isolated.

\subsection{OpenRouter BYOK Mode}

When users supply their own provider credentials via OpenRouter's BYOK (Bring Your Own Key) feature, cache isolation is restored.
Cross-account BYOK tests show no statistical significance (p $>$ 0.16 for all providers) and no metadata-based cache sharing.
This confirms that the vulnerability stems from OpenRouter's use of shared organizational credentials rather than from OpenRouter's routing layer itself.

\begin{figure*}[htbp]
    \centering
    \includegraphics[width=0.8\textwidth]{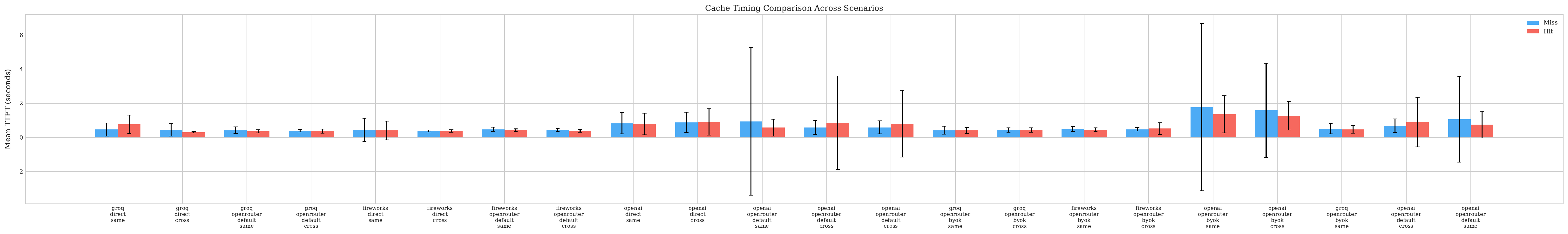}
    \caption{Comparison of cross-account cache detection across routing configurations. Cache sharing is detected via OpenRouter default credentials but not via direct API access or BYOK mode.}
    \label{fig:comparison}
\end{figure*}

\begin{figure*}[htbp]
    \centering
    \includegraphics[width=0.8\textwidth]{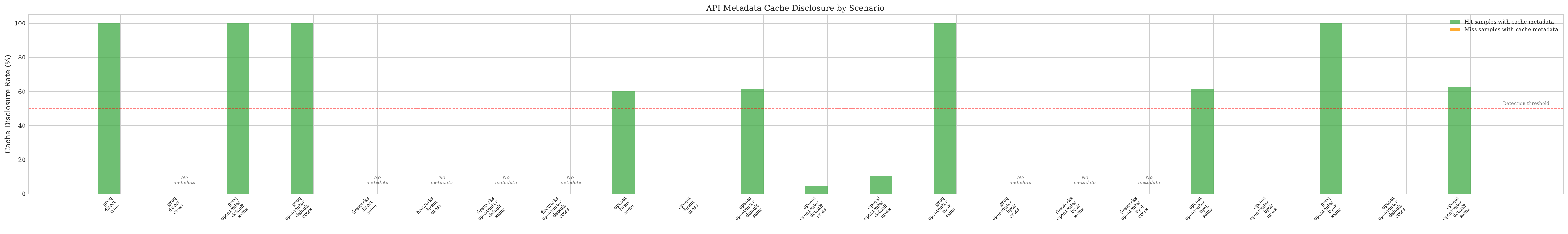}
    \caption{Metadata disclosure rates showing cached token percentages across scenarios.}
    \label{fig:metadata}
\end{figure*}

\section{Discussion}
\label{sec:discussion}

\subsection{Security Implications}

Our findings demonstrate that OpenRouter's default credential-sharing architecture effectively bypasses the per-organization cache isolation that providers have implemented.
This creates a novel attack surface where any OpenRouter user can probe the prompt cache and potentially reconstruct prompts submitted by other users.
The attack is particularly concerning because it requires no special access or privileges beyond a standard OpenRouter account with API credits.

Practical exploitation scenarios include system prompt extraction, where an attacker could reconstruct the system prompts of applications built on OpenRouter, potentially revealing proprietary instructions, RAG retrieval patterns, or confidential business logic.
Additionally, usage pattern detection could allow competitors to determine whether a target organization is using specific prompts or workflows, enabling corporate espionage even without full prompt reconstruction.
The cached prefix information could also reveal sensitive user data embedded in prompts, such as personally identifiable information or proprietary documents.

\subsection{Responsibility Attribution}

Our results reveal a shared responsibility gap between inference providers and API gateways.
Providers have correctly implemented organization-level cache isolation, as evidenced by our direct API baseline tests showing no cross-account cache sharing.
However, this isolation assumes that each organization's API credentials represent a trust boundary.
OpenRouter's architecture violates this assumption by pooling all user traffic under shared organizational credentials.

From the provider's perspective, all OpenRouter traffic originates from a single trusted organization, making cache sharing within that traffic appear intentional.
From OpenRouter's perspective, they are simply proxying requests and may not have visibility into how providers implement caching.
This architectural mismatch creates a vulnerability that neither party may recognize without explicit testing.

While this paper focuses on OpenRouter, the vulnerability pattern is not unique to any single gateway.
Any API aggregation service that proxies requests to upstream providers using shared organizational credentials could inadvertently create cross-user cache sharing.
This includes other LLM API gateways, enterprise API management platforms, and custom proxy implementations.
As the LLM ecosystem increasingly relies on intermediary services for routing, cost optimization, and unified API access, the cache isolation implications of shared credentials deserve broader attention.

\subsection{Mitigations}

The most robust mitigation would require providers to implement true per-tenant cache namespace isolation, where cache entries are explicitly partitioned by a tenant identifier rather than shared across an organization.
OpenRouter could then forward a unique per-user identifier with each request, enabling providers to isolate caches by OpenRouter user even when all requests arrive under shared organizational credentials.
This approach would preserve response quality and caching benefits while restoring the isolation guarantees that users expect.

Some existing mechanisms offer partial mitigation.
OpenAI's \texttt{prompt\_cache\_key} parameter influences which cache server handles a request, so using distinct keys for different users would probabilistically reduce cache sharing by routing users to different servers.
However, this is a routing hint rather than a true cache namespace; with a finite number of cache servers, collisions remain possible and isolation is not guaranteed.

Users concerned about prompt privacy should use OpenRouter's BYOK (Bring Your Own Key) mode, which our results confirm restores proper cache isolation.
However, this sacrifices the convenience and pricing benefits that make OpenRouter attractive in the first place.

\subsection{Limitations}

Our study has several limitations that should be considered when interpreting the results.
First, we tested only three providers and one model per provider; cache isolation behavior may vary across other providers or models available through OpenRouter.
Second, our experiments used randomly generated prompts rather than realistic application prompts, which may not reflect typical cache hit patterns in production.
Third, cache eviction policies and timing may vary based on provider load, time of day, or other factors we did not control for.
Finally, our statistical thresholds, while conservative, may not capture all forms of cache leakage, particularly in cases of partial or intermittent sharing.

\section*{Responsible Disclosure}
We initially contacted OpenRouter regarding these findings on November 14th, 2025, and followed up on December 12th, 2025 with a draft of this paper. 
OpenRouter acknowledged the report and requested access to our GitHub repository, which was private at the time.
We sent an invitation to OpenRouter's team but did not receive a response.

\section{Conclusion}
\label{sec:conclusion}

This paper investigated whether OpenRouter's API gateway architecture introduces prompt caching vulnerabilities that bypass provider-level isolation guarantees.
Through systematic testing across three major inference providers, we demonstrated that OpenRouter's use of shared organizational credentials enables cross-account cache sharing that would not occur when accessing providers directly.
All three providers tested exhibited detectable cache sharing in OpenRouter's default mode, while properly isolating caches when accessed via direct API or OpenRouter's BYOK mode.

These findings have immediate practical implications.
OpenRouter users should be aware that their prompts may be observable by other users through timing side-channels and metadata disclosure.
Organizations handling sensitive data should consider using BYOK mode or direct provider access to maintain cache isolation.
We recommend that OpenRouter implement per-user cache isolation mechanisms, and that providers consider offering more granular isolation options to better support multi-tenant gateway architectures.

\onecolumn
\appendices
\section{Data Availability}
\label{sec:data}

The complete source code for CacheProbe, including all experimental results and raw timing data, is publicly available at \url{https://github.com/Ryan5453/cacheprobe}.

\section{Detailed Timing Distributions}
\label{sec:timing-appendix}

\begin{figure}[H]
    \centering
    \includegraphics[width=0.30\textwidth]{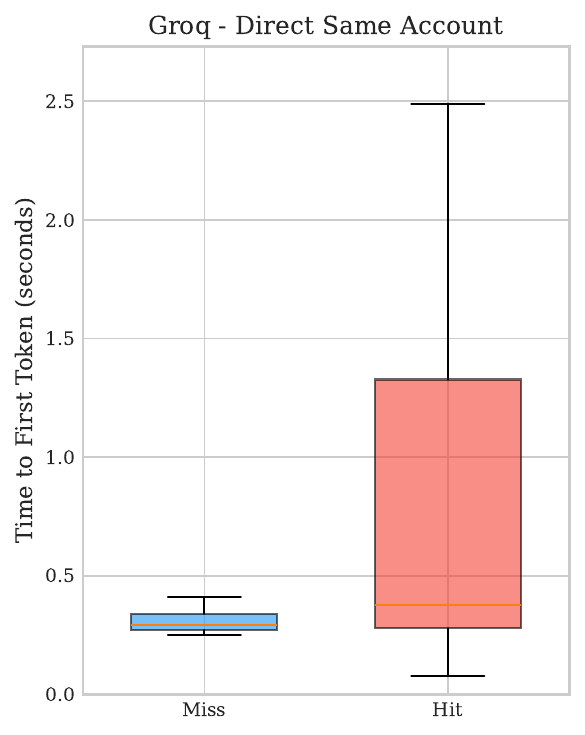}
    \includegraphics[width=0.30\textwidth]{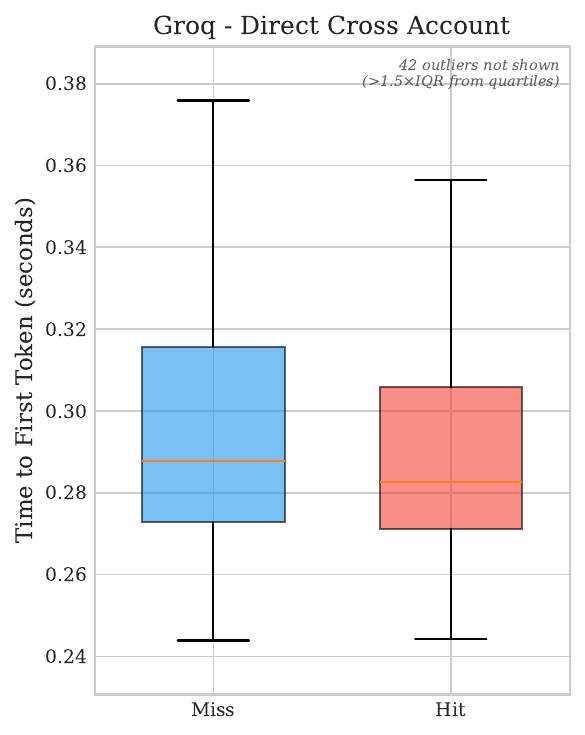}
    \includegraphics[width=0.30\textwidth]{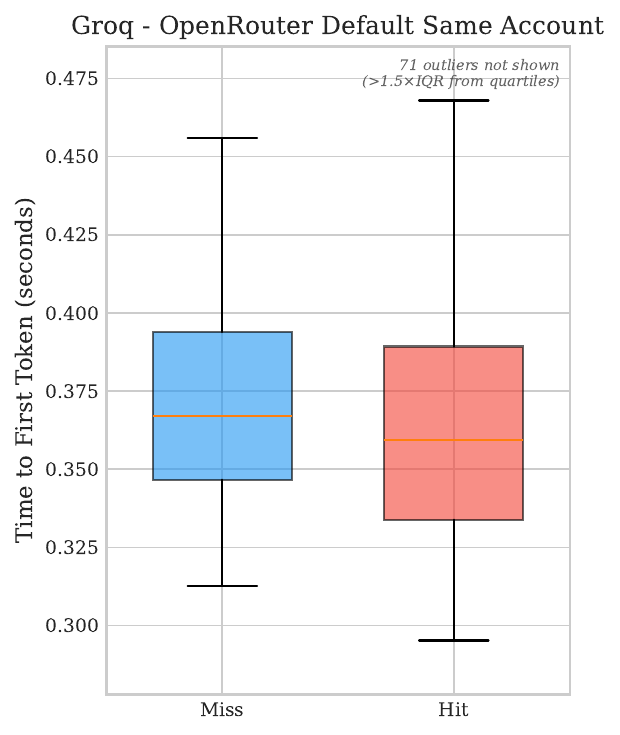}
    \includegraphics[width=0.30\textwidth]{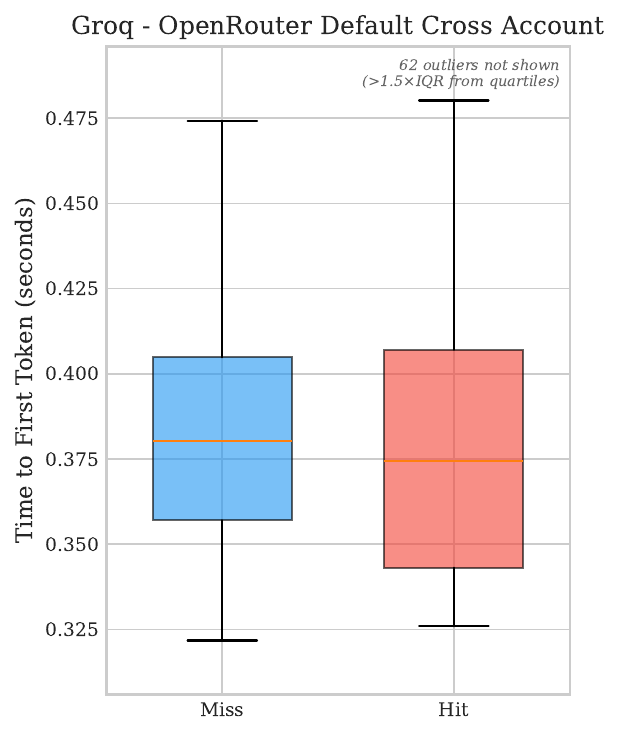}
    \includegraphics[width=0.30\textwidth]{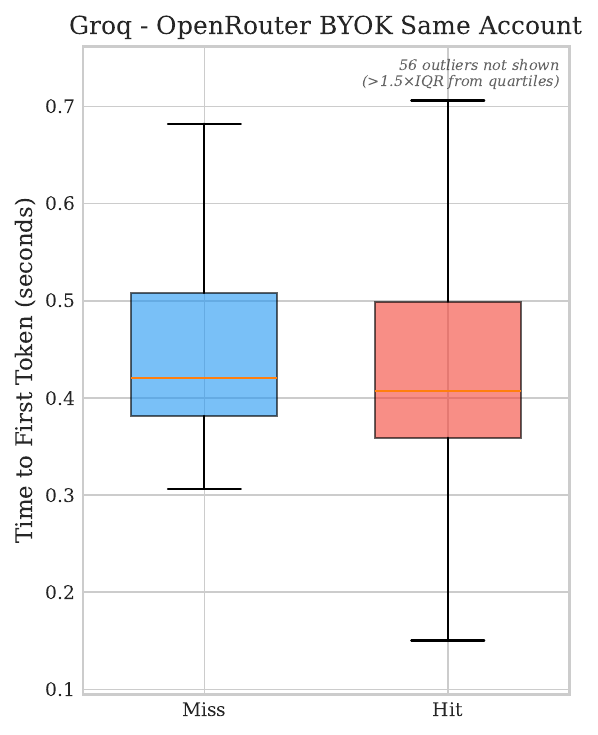}
    \includegraphics[width=0.30\textwidth]{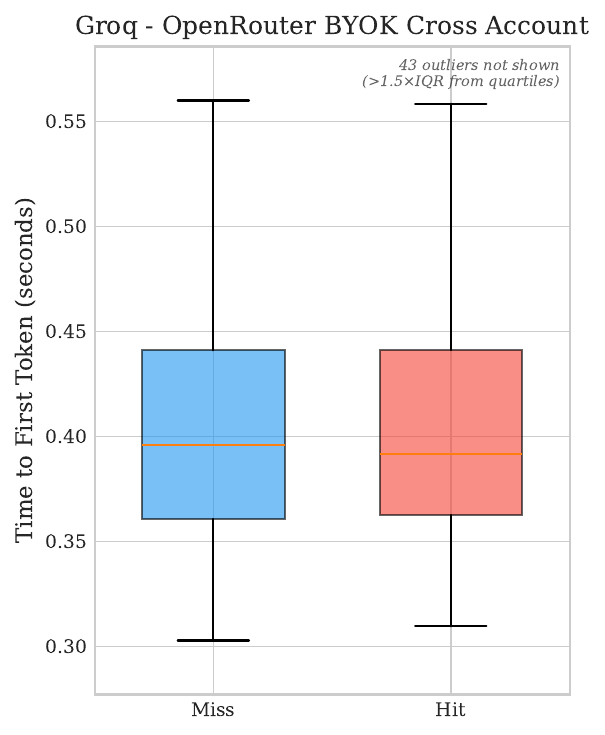}
    \caption{Groq TTFT boxplots across all scenarios.}
    \label{fig:groq-boxplots}
\end{figure}

\begin{figure}[H]
    \centering
    \includegraphics[width=0.45\textwidth]{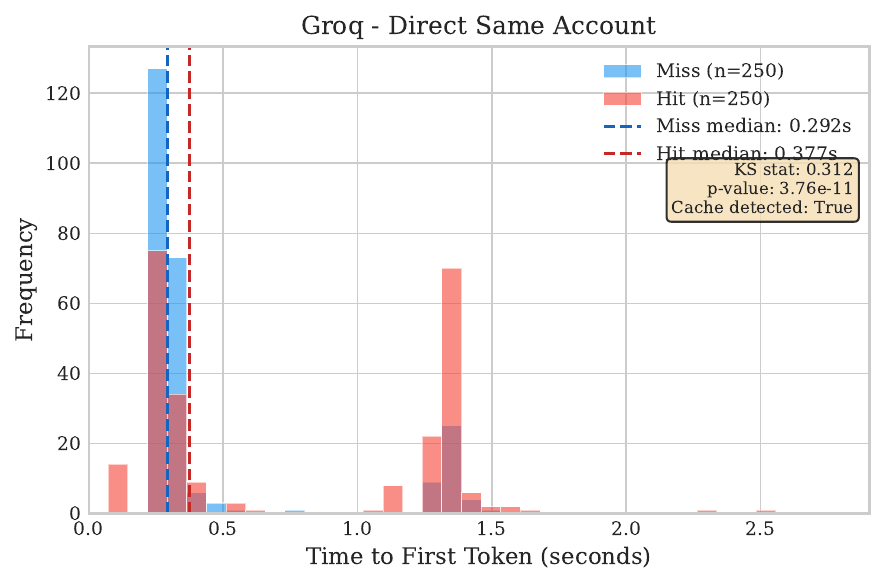}
    \includegraphics[width=0.45\textwidth]{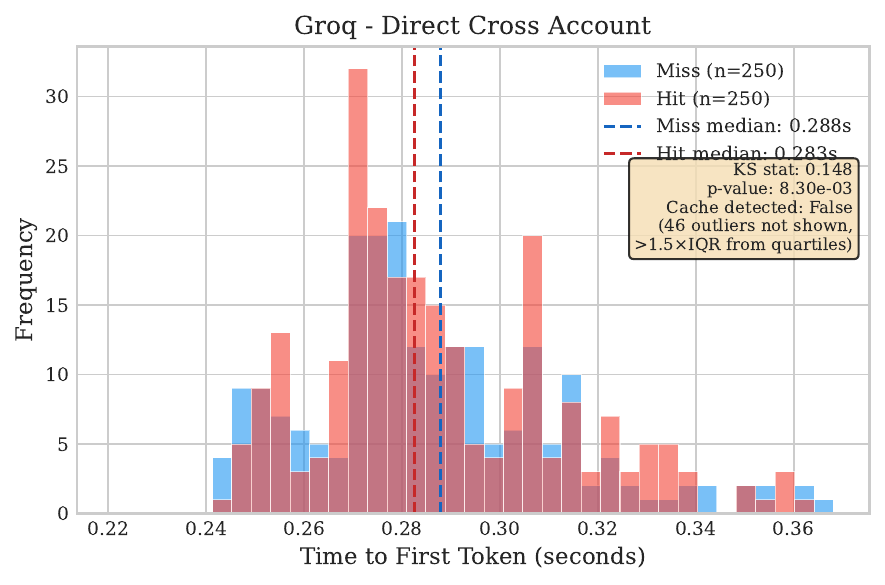}
    \includegraphics[width=0.45\textwidth]{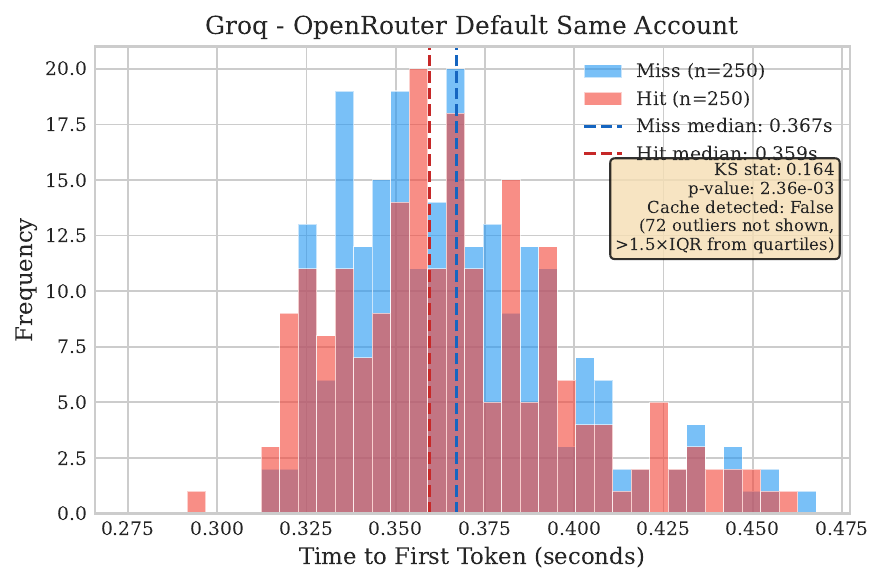}
    \includegraphics[width=0.45\textwidth]{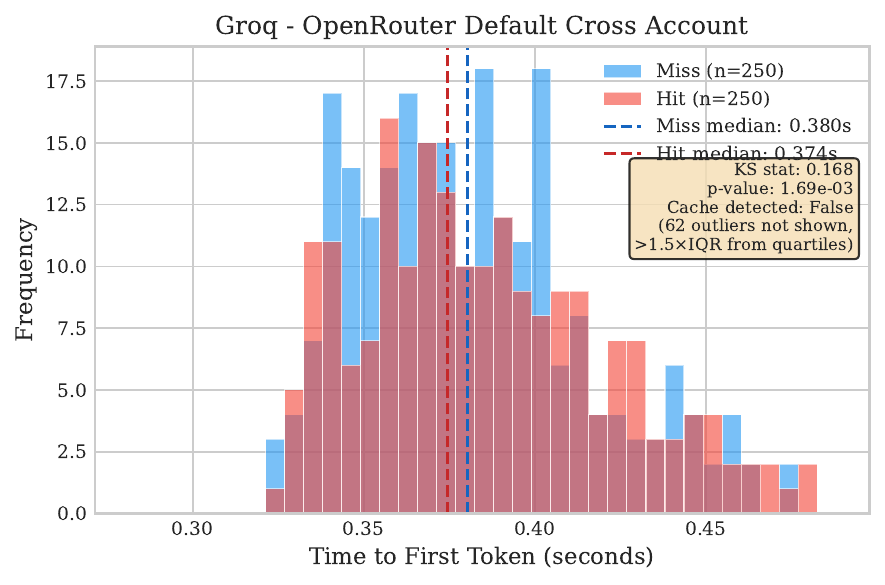}
    \includegraphics[width=0.45\textwidth]{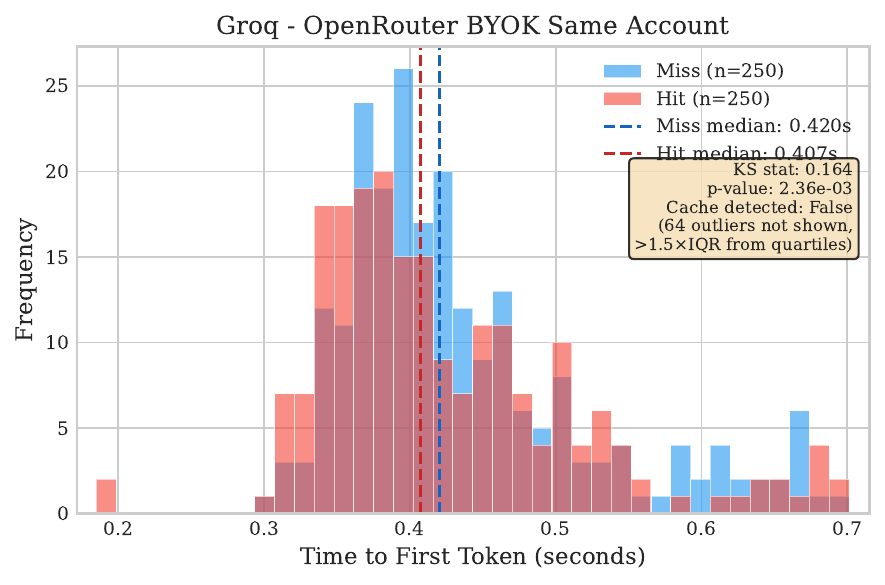}
    \includegraphics[width=0.45\textwidth]{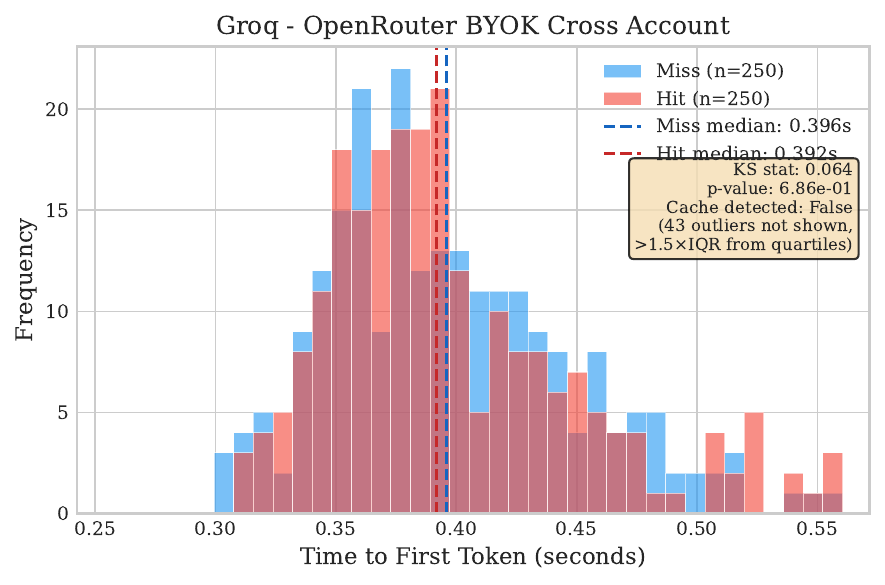}
    \caption{Groq TTFT histograms across all scenarios.}
    \label{fig:groq-histograms}
\end{figure}

\begin{figure}[H]
    \centering
    \includegraphics[width=0.30\textwidth]{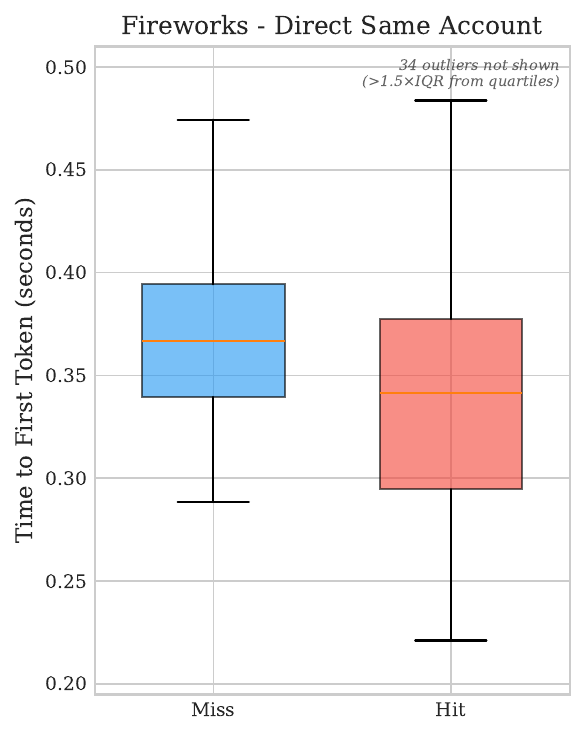}
    \includegraphics[width=0.30\textwidth]{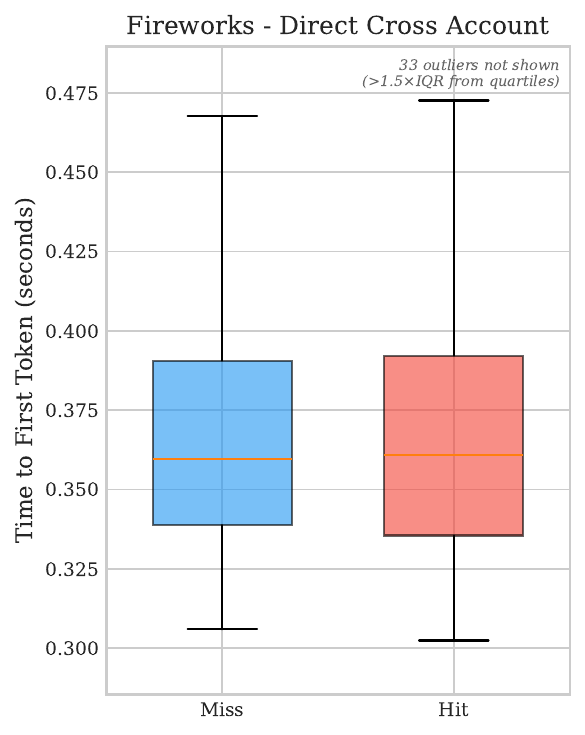}
    \includegraphics[width=0.30\textwidth]{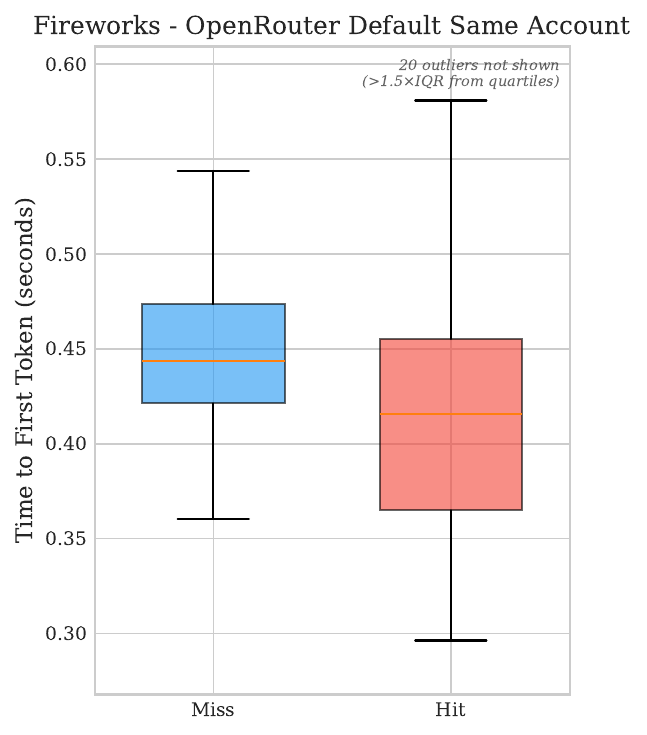}
    \includegraphics[width=0.30\textwidth]{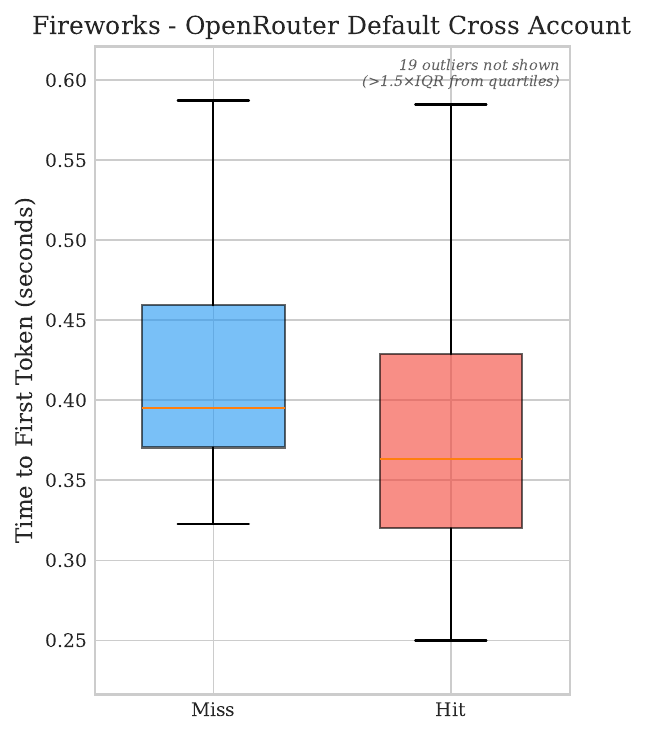}
    \includegraphics[width=0.30\textwidth]{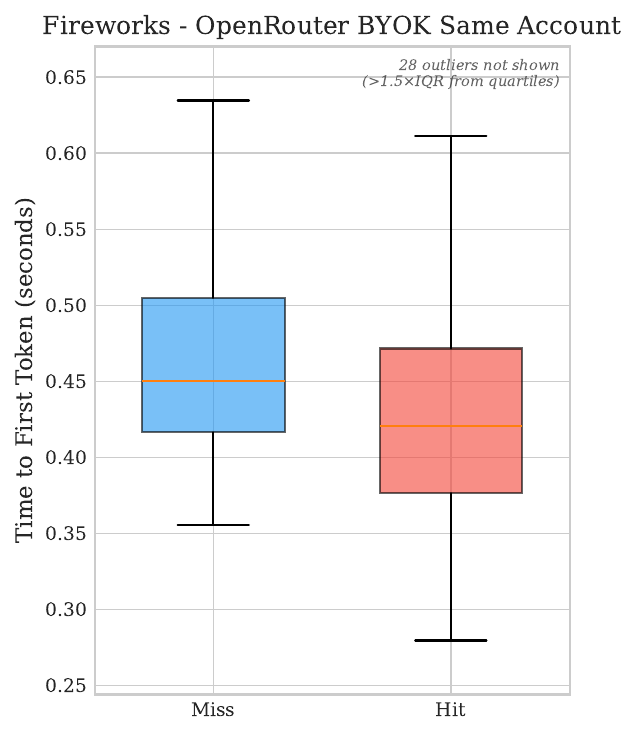}
    \includegraphics[width=0.30\textwidth]{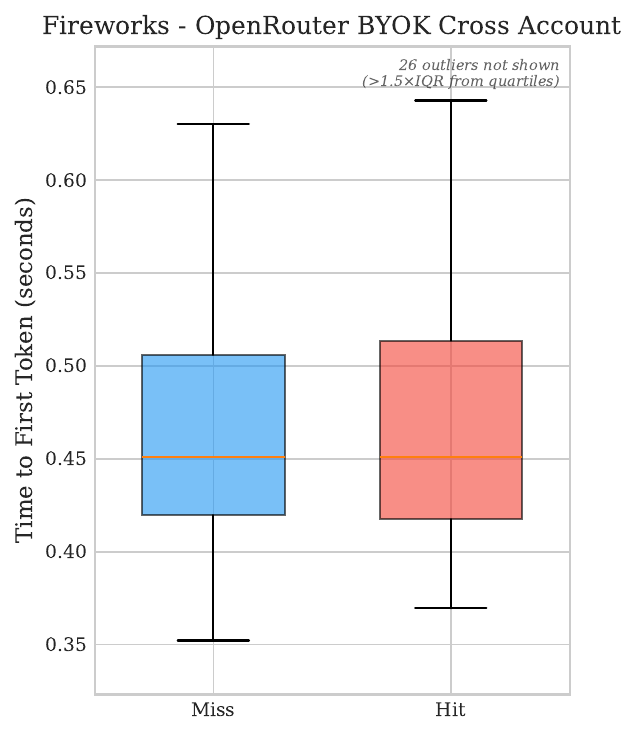}
    \caption{Fireworks TTFT boxplots across all scenarios.}
    \label{fig:fireworks-boxplots}
\end{figure}

\begin{figure}[H]
    \centering
    \includegraphics[width=0.45\textwidth]{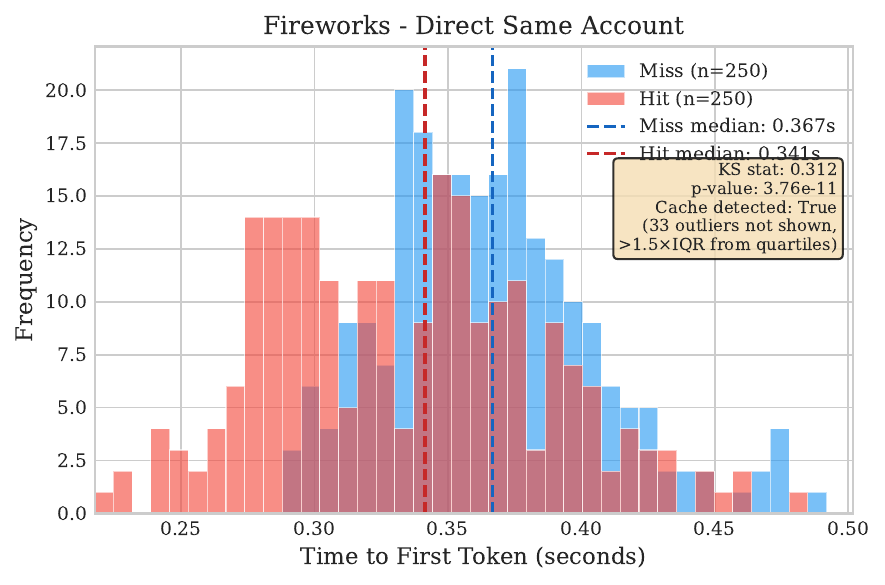}
    \includegraphics[width=0.45\textwidth]{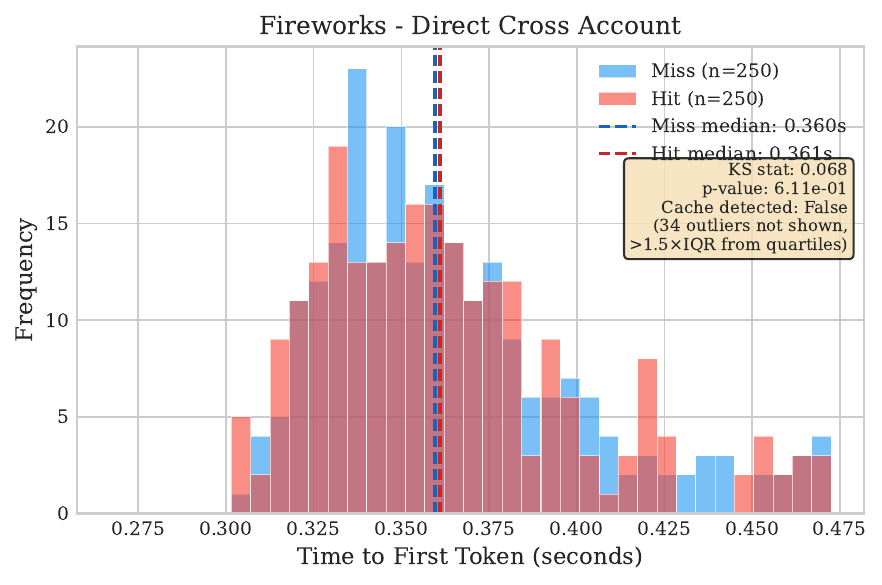}
    \includegraphics[width=0.45\textwidth]{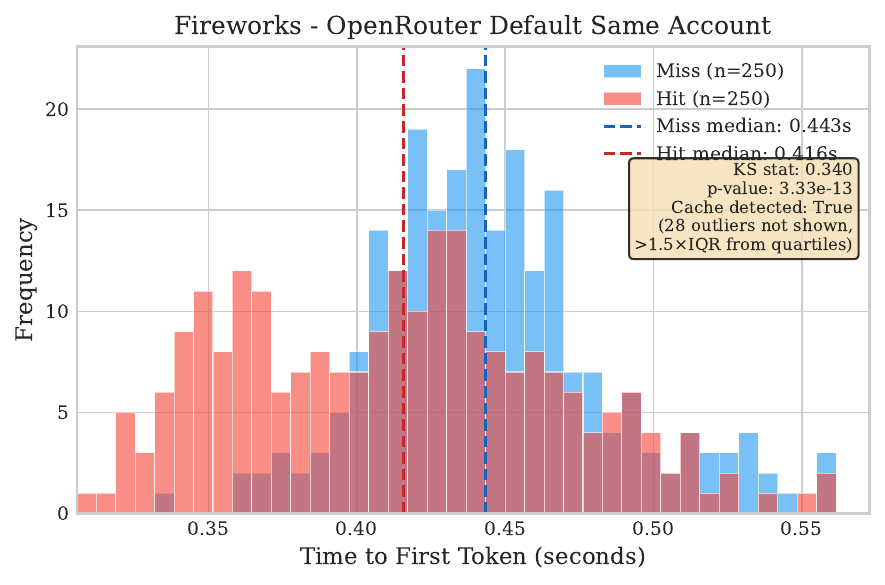}
    \includegraphics[width=0.45\textwidth]{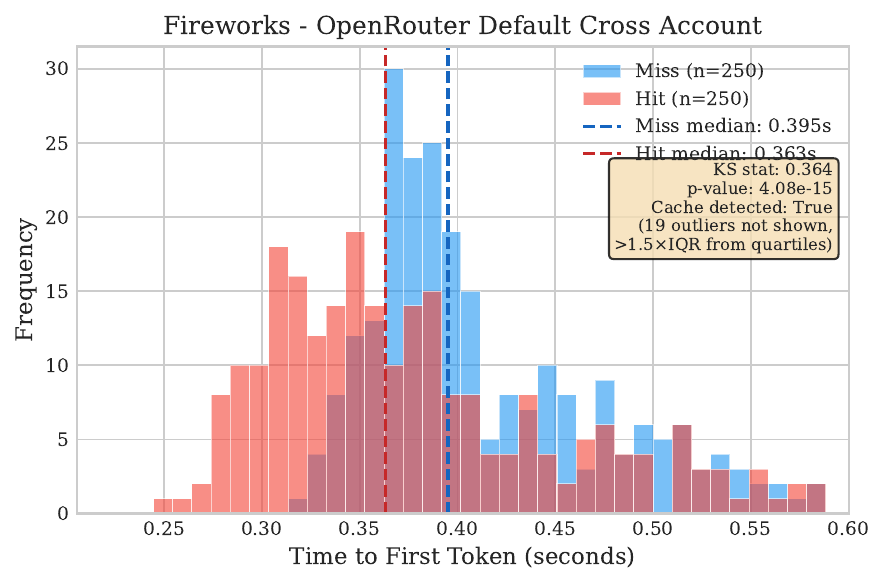}
    \includegraphics[width=0.45\textwidth]{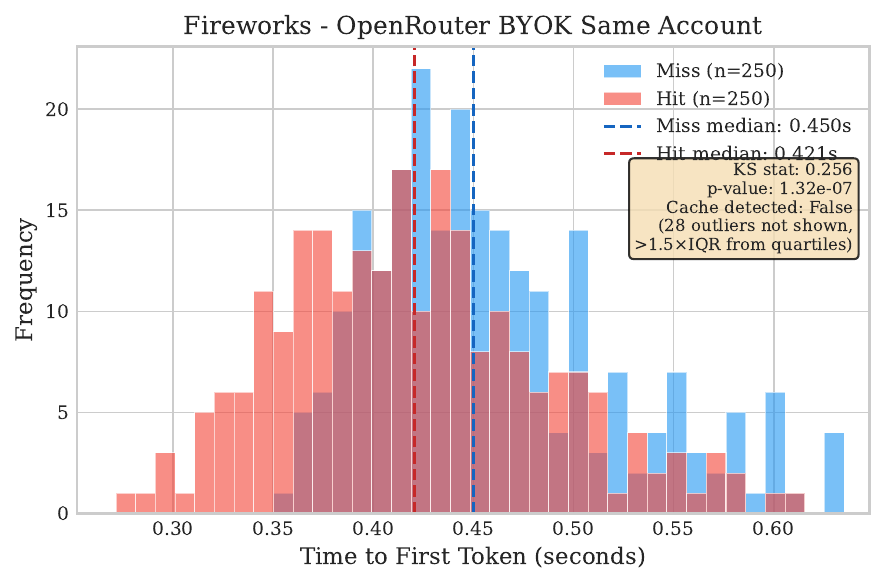}
    \includegraphics[width=0.45\textwidth]{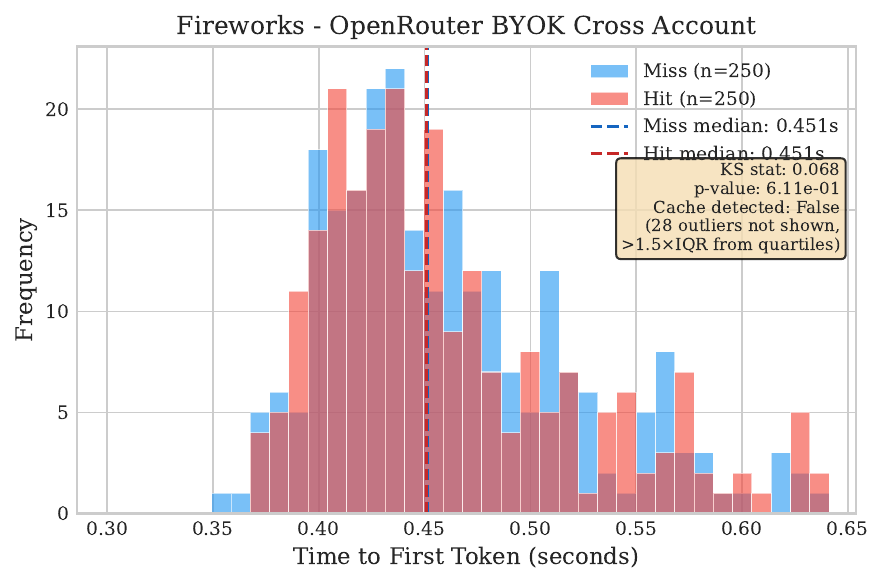}
    \caption{Fireworks TTFT histograms across all scenarios.}
    \label{fig:fireworks-histograms}
\end{figure}

\begin{figure}[H]
    \centering
    \includegraphics[width=0.30\textwidth]{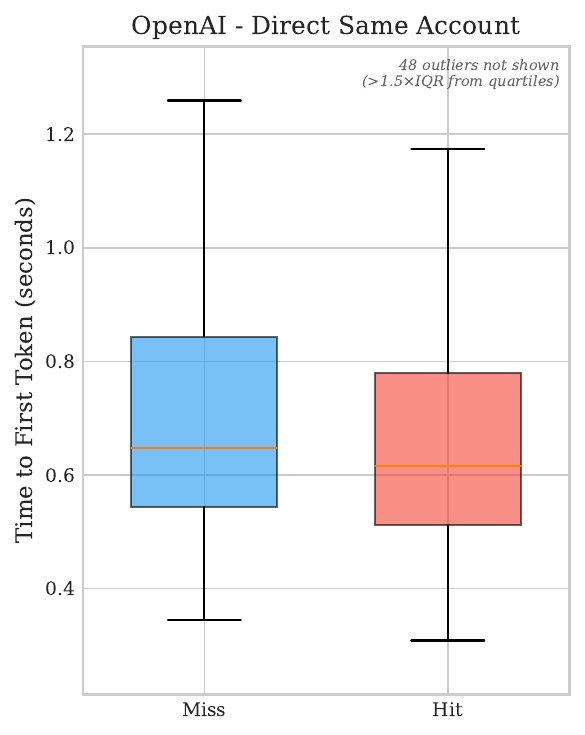}
    \includegraphics[width=0.30\textwidth]{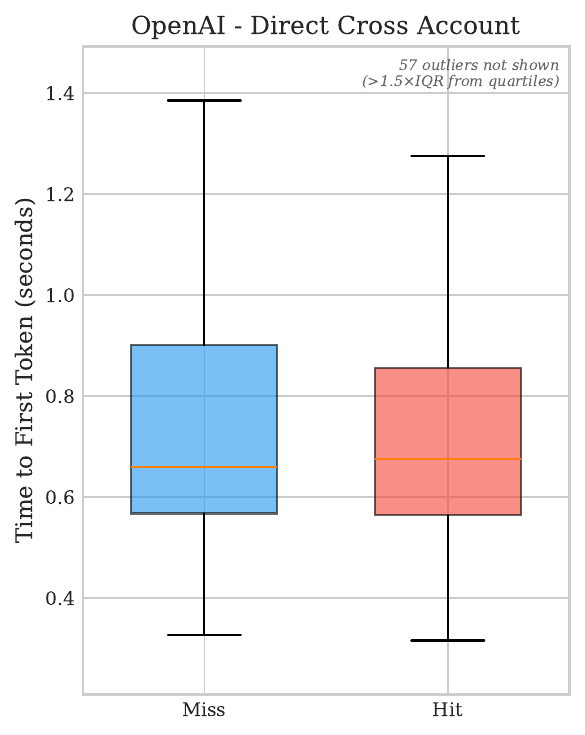}
    \includegraphics[width=0.30\textwidth]{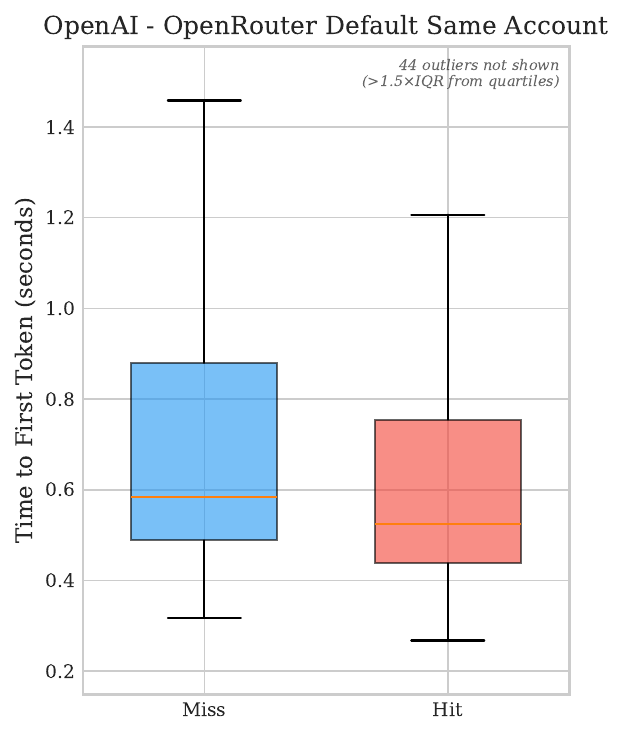}
    \includegraphics[width=0.30\textwidth]{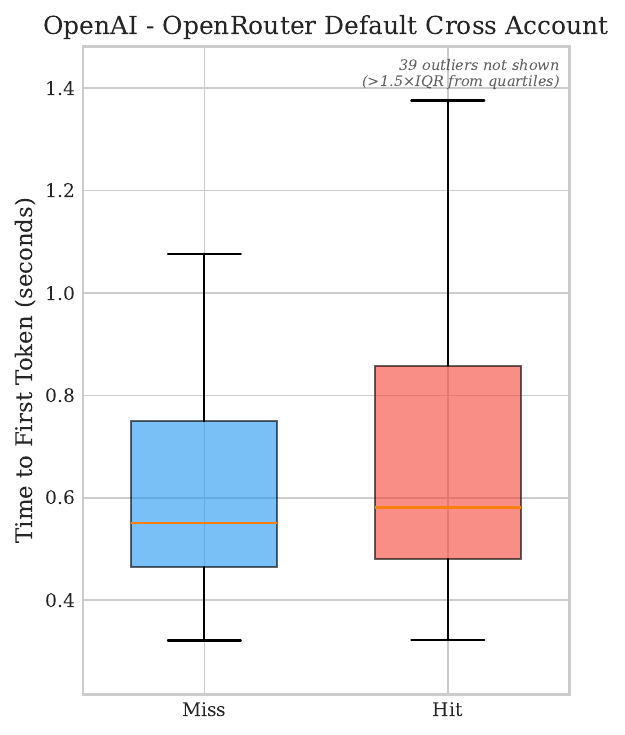}
    \includegraphics[width=0.30\textwidth]{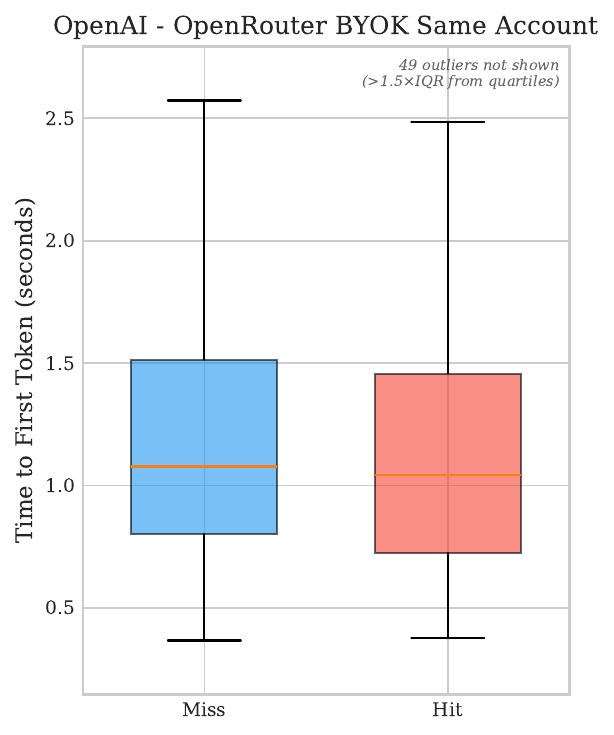}
    \includegraphics[width=0.30\textwidth]{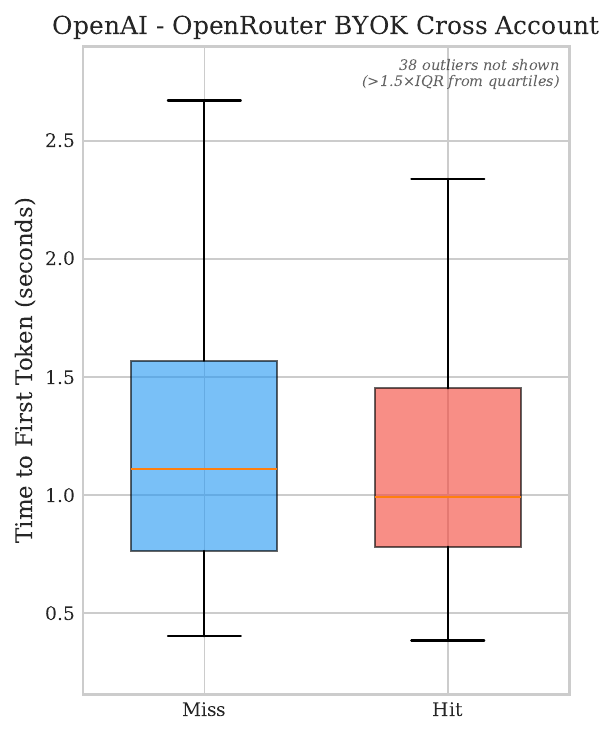}
    \caption{OpenAI TTFT boxplots across all scenarios.}
    \label{fig:openai-boxplots}
\end{figure}

\begin{figure}[H]
    \centering
    \includegraphics[width=0.45\textwidth]{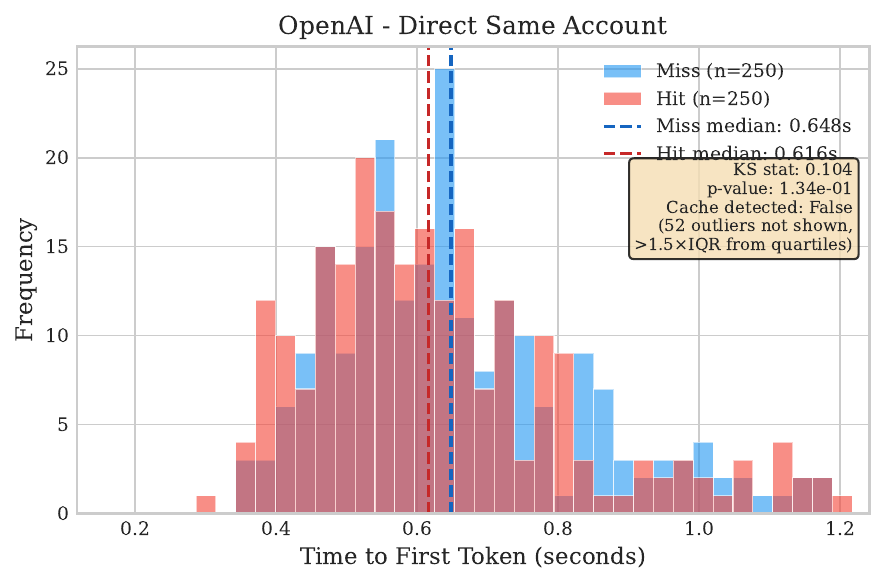}
    \includegraphics[width=0.45\textwidth]{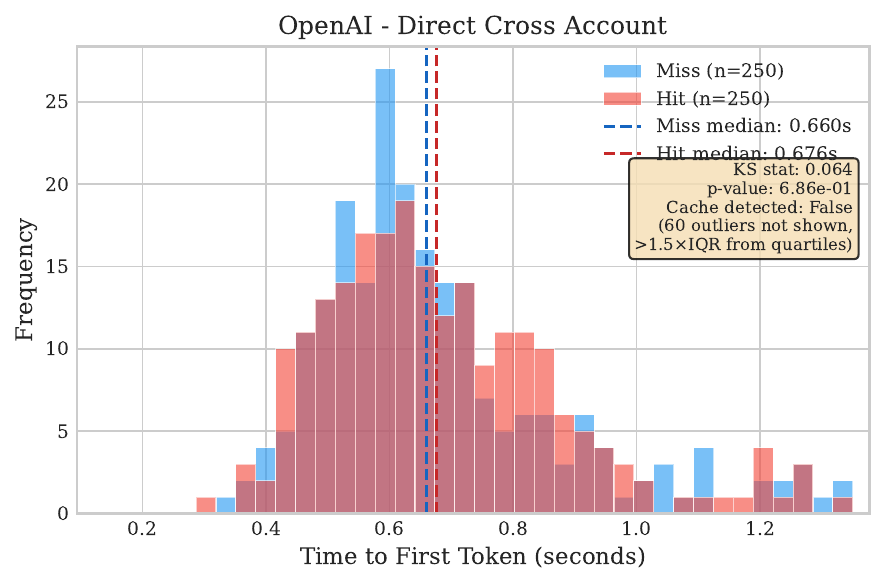}
    \includegraphics[width=0.45\textwidth]{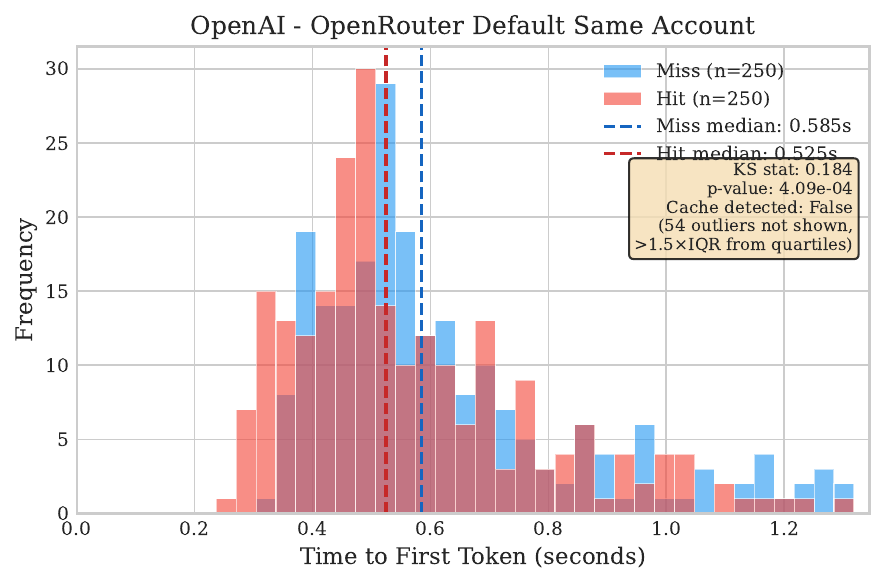}
    \includegraphics[width=0.45\textwidth]{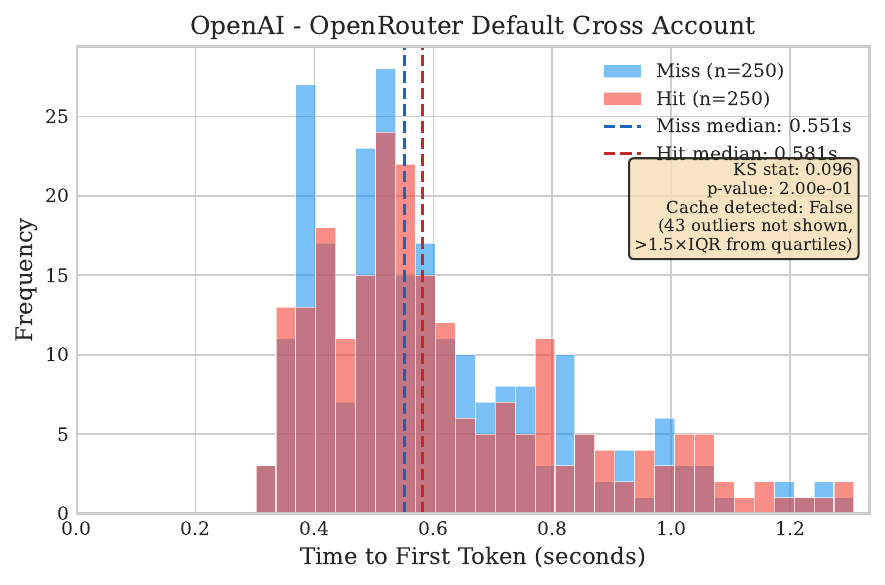}
    \includegraphics[width=0.45\textwidth]{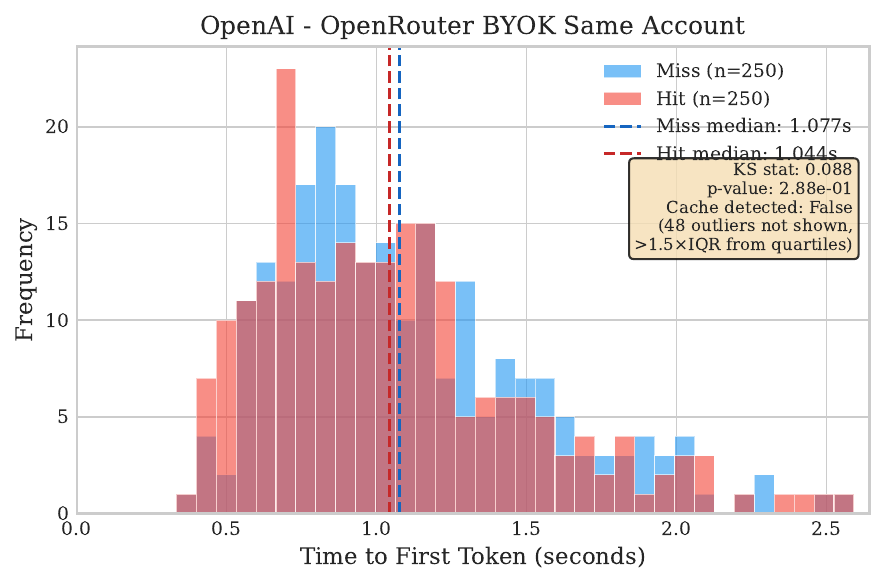}
    \includegraphics[width=0.45\textwidth]{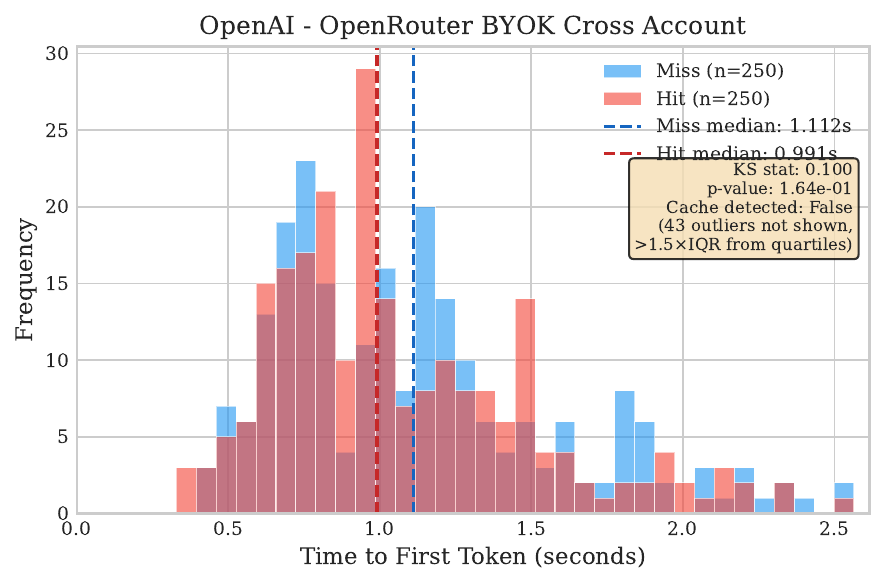}
    \caption{OpenAI TTFT histograms across all scenarios.}
    \label{fig:openai-histograms}
\end{figure}

\bibliographystyle{IEEEtran}
\bibliography{references}

\end{document}